\documentclass[11pt]{article}
\usepackage[normalem]{ulem}
\pdfoutput=1 % if your are submitting a pdflatex (i.e. if you have

\usepackage{jheppub} % for details on the use of the package, please
\usepackage{url}
\usepackage{caption}
\usepackage{subcaption}

\usepackage[latin9]{inputenc}
\usepackage{float}
\usepackage{amsmath}
\usepackage{amssymb}
\usepackage{graphicx}
\usepackage{hyperref}
\usepackage{color}
\usepackage{bbm}
\newcommand{\be}{\begin{equation}}
\newcommand{\ee}{\end{equation}}
\newcommand{\ben}{\begin{displaymath}}
\newcommand{\een}{\end{displaymath}}
\newcommand{\bea}{\begin{eqnarray}}
\newcommand{\eea}{\end{eqnarray}}
\def\K{K{\"a}hler }
   \newcommand{\rf}[1]{(\ref{#1})}
\newcommand{\vp}{\varphi}

\def\be{\begin{equation}}
\def\ee{\end{equation}}
\def\bea{\begin{eqnarray}}
\def\eea{\end{eqnarray}}
\def\ba{\begin{array}}
\def\ea{\end{array}}
\def\bit{\begin{itemize}}
\def\eit{\end{itemize}}

\def\a{\alpha}

\def\vp{\varphi}

\def\vt{\vartheta}

 \makeatletter

\allowdisplaybreaks

\makeatother

\makeatletter
\DeclareRobustCommand{\rcite}[1]{%
  \rcite@aux#1,\@nil{#1}%
}
\def\rcite@aux#1,#2\@nil#3{%
  \if\relax#2\relax
    Ref.~\cite{#3}%
  \else
    Refs.~\cite{#3}%
  \fi
}
\makeatother

\hypersetup{
    colorlinks = true,
    citecolor = {blue},
    linkcolor = {blue},
    urlcolor = {blue},
}

 \title{\rm {\bf \LARGE \boldmath  { Axion Stabilization in Modular Cosmology}}}

\author{John Joseph Carrasco$^1$,}
\author{Renata Kallosh$^2$, }
\author{Andrei Linde$^2$ and}
\author{Diederik Roest$^3$}

\affiliation{1:Amplitudes and Insights Group, Department of Physics and Astronomy,\\ Northwestern University, Evanston, IL 60208, USA}
\affiliation{2: Stanford Institute for Theoretical Physics and Department of Physics,\\ Stanford University, Stanford, CA 94305, USA}
\affiliation{3: Van Swinderen Institute for Particle Physics and Gravity, University of Groningen, \\ 9747 AG Groningen, The Netherlands}
\emailAdd{carrasco@northwestern.edu}
\emailAdd{kallosh@stanford.edu}
\emailAdd{alinde@stanford.edu}
\emailAdd{d.roest@rug.nl}

\abstract{  
 The $SL(2,\mathbb{Z})$ invariant $\alpha$-attractor models proposed in \cite{Kallosh:2024ymt} have plateau potentials with respect to the inflaton and axion fields. The potential in the axion direction is almost exactly flat during inflation, hence the axion field remains nearly massless.  In this paper, we develop a generalized class of such models, where the $SL(2,\mathbb{Z})$ symmetry is preserved, but the axion acquires a large mass and becomes strongly stabilized during inflation, which eliminates isocurvature perturbations in this scenario. Inflation in such two-field models occurs as in the single-field $\alpha$-attractors and leads to the same cosmological predictions.

}

\begin{document}

\maketitle

% \tableofcontents{}

%\newpage

\section{Introduction}\label{Sec:intro}

\parskip 7.5pt

Cosmological models with a natural embedding into supergravity or superstring theory often feature multiple scalar fields. A prominent example is that of $\alpha$-attractors \cite{Kallosh:2013yoa}, which crucially relies on the hyperbolic geometry of the scalar manifold spanned by the axion-dilaton field. In addition to the inflaton and the corresponding curvature fluctuations, such multi-field models feature additional scalar directions that can lead to the (possibly copious) production of isocurvature fluctuations. 

In order to be compatible with existing CMB constraints on non-adiabatic modes \cite{Planck:2018jri}, the simplest approach is to stabilize the additional field(s) by making them massive, with masses greater than the Hubble constant during inflation. In this way, one can bring these models to the theory of a single-field inflation. A detailed analysis of such models was performed in the context of $\a$-attractor models with a stabilized axion field in \cite{Kallosh:2013yoa, Carrasco:2015uma, Carrasco:2015rva, Kallosh:2017wnt}.

An interesting perspective was developed very recently by augmenting the $SL(2,\mathbb{R})$-invariant hyperbolic geometry with a scalar potential that preserves a discrete $SL(2,\mathbb{Z})$ subgroup \cite{Casas:2024jbw,Kallosh:2024ymt,Kallosh:2024pat,Kallosh:2024whb}, referred to as modular cosmology. 
These models are inspired by string theory, see in particular  \cite{Casas:2024jbw}, so it is important to develop their supergravity generalization without breaking  $SL(2,\mathbb{Z})$ invariance. This problem was solved in \cite{Kallosh:2024ymt} without changing any properties of the bosonic part of the action. 

One of these properties is the extraordinary flatness of the inflationary potential in the axion direction: it is doubly exponentially suppressed at large inflaton values \cite{Kallosh:2024whb}. This results in large isocurvature perturbations of the axion field. One can show that during inflation, these perturbations do not destabilize the inflationary trajectory and do not feed into curvature perturbations. However, there is an unconventional possibility that these isocurvature perturbations may feed into the large-scale curvature perturbations {\it after inflation}  \cite{Kallosh:2024whb}. This scenario requires a detailed investigation and will be discussed in a separate publication \cite{6authors}. 

In the current paper, we will investigate another option outlined above: that of axion stabilization. A priori, it is not clear that one can achieve this in an $SL(2,\mathbb{Z})$-manner. All previously developed methods for stabilization involve the introduction of a $(\tau+ \bar\tau)^{2}$ term in the potential, such that the field $\tau+ \bar\tau$ acquires a mass and quickly settles down to $\tau+ \bar\tau=0$. This was done using a heavy stabilized field in general supergravity models in \cite{Kawasaki:2000yn,Kallosh:2010ug,Kallosh:2010xz}, and using a nilpotent stabilizer in $\a$-attractor models \cite{Carrasco:2015uma,Carrasco:2015rva,Kallosh:2017wnt}. However, the introduction of a $(\tau+ \bar\tau)^{2}$  term would break the $SL(2,\mathbb{Z})$ invariance of the potential. 

 We will demonstrate in this paper that one can stabilize the axion in an $SL(2,\mathbb{Z})$-compatible manner in the models based on $j$-functions  \cite{Kallosh:2024ymt}. Note that $j$-function and its complex conjugate $\overline {j(\tau)}$ are $SL(2,\mathbb{Z})$ invariant, so any function of $j$ and $\overline {j(\tau)}$ is $SL(2,\mathbb{Z})$ invariant. This allows to introduce axion stabilization via terms such as  $|j(\tau) \pm \overline {j(\tau)}|^{2}$, or more general terms, such as  $|j(\tau)\, e^{-i 2\pi \gamma} - \overline {j(\tau)} \, e^{i 2\pi \gamma}|^{2}$ where $\gamma$ is some constant. As we will see, this leads to the formation of axion valleys, stabilizing the axion field. Inflation in such two-field models typically occurs in the same way as in the single-field $\alpha$-attractors, and therefore leads to the same cosmological predictions. 

%It is relatively easy to find an $SL(2,\mathbb{Z})$ invariant axion stabilization term depending on $j(\tau)$-function because the  $j$-function and its complex conjugate $\overline {j(\tau)}$ are $SL(2,\mathbb{Z})$ invariant, so any function of $j$ and $\overline {j(\tau)}$ is $SL(2,\mathbb{Z})$ invariant. 

 In comparison, other models described in  \cite{Kallosh:2024ymt} have $SL(2,\mathbb{Z})$ invariant potentials depending on a combination of modular forms, which transform under $SL(2,\mathbb{Z})$, and on a complex field $\tau$.  These involve a Dedekind function $\eta(\tau)$, a modular form of weight 1/2, an almost holomorphic modular form of weight 2, $\tilde G_2$ etc. It is not immediately clear how to construct $SL(2,\mathbb{Z})$ invariant axion stabilization terms in these models; see  \cite{Kallosh:2024whb,6authors} for an investigation of the cosmological consequences of the models without axion stabilization.

The stabilization procedure is described in Secs. \ref{Sec:stab}-\ref{Sec:Kil} and summarized in Sec. \ref{Sec:sum}. We present general analytic arguments why  slow-roll inflationary trajectories in these models quickly reach the axion valleys and proceed to the minima of the potentials along the valleys, as in single-field inflationary models. We also give examples of various inflationary trajectories illustrating the general properties of the stabilized $SL(2,\mathbb{Z})$ models.  The supergravity version of these stabilized $SL(2,\mathbb{Z})$ invariant models involving a nilpotent stabilizer superfield is presented in Appendix \ref{App:A}. It is valid for any value of the parameter $\a$ in $\a$-attractor models. In Appendix \ref{App:B} we describe different minima of these models,  reachable at the end of inflation: one is at the boundary of the fundamental domain,  while the other does not border the fundamental domain but lies an $SL(2,\mathbb{Z})$ transformation away from it. 

\section{Axion stabilization preserving $SL(2,\mathbb{Z})$ invariance}\label{Sec:stab}

Our starting point will be the hyperbolic geometry spanned by the axion-dilaton field of $\alpha$-attractors:
 \begin{align}
  ds^2 = 6 \alpha \frac{d \tau d \bar \tau}{|\tau - \bar \tau|^2} = d \varphi^2  + \frac{3\alpha}{2} e^{-2\sqrt{\frac{2}{3\alpha}}\vp} d \theta^2 \,, \qquad \tau = \theta + i e^{\sqrt{2\over 3\alpha} \varphi} \,, \label{hyperbolic}
 \end{align}
 with $SL(2,\mathbb{R})$ isometry group, setting the kinetic terms for the axion and dilaton field. A subset of the current authors has proposed  \cite{Kallosh:2024ymt} to introduce a scalar potential in a manner compatible with the discrete $SL(2,\mathbb{Z})$ subgroup of the hyperbolic isometry group. These models are naturally formulated in terms of modular functions. A prime example is given by 
\be
V_\beta =V_0\Big (1-{\ln \beta^{2}
\over  \ln \big(|j(\tau)|^2 + \beta^{2}\big)}\Big ) \ .
\label{beta1}\ee
in terms of the $j$-function and with free parameter $\beta$. 

A natural choice for this parameter is $\beta = j(i) = 12^3$ due to the relation of the $j$-function to Klein's Absolute invariant $J$, given by $j = 12^3 J$ and with normalization $J(i)=1$.  For this parameter choice, the potential takes the esthetically pleasing form
\be
V =V_0\Big (1-{\ln |j(i)|^2
\over  \ln \big(|j(\tau)|^2 + |j(i)|^2\big)}\Big ) \ ,
\label{beta}\ee
and is solely defined in terms of the $j$-function.

It was shown in \cite{Kallosh:2024whb} that the $\theta$-derivatives of the inflationary potential are double exponentially suppressed during inflation, so the axion field during inflation is almost exactly flat. As we mentioned in the Introduction, this leads to isocurvature perturbations, which, in certain cases, may affect the cosmological predictions of these models.

In the present paper, we will outline a number of possibilities to avoid this potential problem by making the axion massive and stabilizing its value during inflation. As a simple example of a model with axion stabilization, consider a potential 
\be
V_\beta^{j-\bar j} =V_0\Big (1-{\ln \beta^{2}
\over  \ln\Big[|j(\tau)|^2+A\,
| j(\tau) -\overline {j(\tau)}|^{2} + \beta^{2}\Big]}\Big ) \ .
\label{beta2}\ee
We will refer to the term with coefficient $A$ (which we assume to be positive) as the stabilization term, for reasons that will become apparent. This potential is $SL(2,\mathbb{Z})$ invariant because $ j(\tau)$ and $\overline {j(\tau)}$ are $SL(2,\mathbb{Z})$ invariants. The supergravity version of the theory is presented in  Appendix A.  

To investigate the properties of this potential during inflation, we note that  $j(\tau)$ at large $\varphi$ can be represented as an expansion
\be
j(\tau)   =   e^{2\pi e^{\sqrt{2\over 3\alpha} \varphi} -2\pi i \theta} +  c_0  +\sum_{n=1} c_n e^{- 2n \pi e^{\sqrt{2\over 3\alpha} \varphi} +2n \pi i \theta} \,.
\label{largej2}\ee
One can see that $j(\tau)$ is real for $\theta = k/2$, where $k $ is any integer. Therefore, in each of these cases, $ j(\tau) -\overline {j(\tau)}= 0$, and the potential at $\theta = k/2$ coincides with the original potential \rf{beta1}. At all other values $\theta \neq k/2$, the stabilization term gives a positive contribution to the potential: 
 \begin{align}
  | j(\tau) -\overline {j(\tau)}|^{2} \to 4 e^{4\pi e^{\sqrt{2\over 3\alpha} \varphi}}\   \sin^{2} (2\pi \theta) \,.
 \end{align}
As a result, the lines $\theta = k/2$ at large $\vp$ correspond to the positions of the axion valleys of the potential. These valleys exist for all values of $A >0$. They become more pronounced as we increase $A$, but as we are going to see, the axion stabilization occurs even for rather small values of  $A$.

For a more detailed investigation of inflation, it is illustrative to study the explicit expansion of the scalar potential at large $\vp$. For the case \eqref{beta1} without stabilization, this reads
 \begin{align}
 V_\beta = V_0 \left( 1 - \frac{\ln \beta}{2\pi} e^{- \sqrt{2\over 3\alpha} \varphi} + \frac{c_0 \ln \beta }{4 \pi^2} e^{- 2\sqrt{2\over 3\alpha} \varphi} e^{-2\pi e^{\sqrt{2\over 3\alpha} \varphi}} \cos(2 \pi \theta) + \ldots 
\right) \,.  
 \end{align}
Note that the dependence of $|j(\tau)^2|$ on the axion $x \equiv  \theta$ appeared only when we took into account the second, subleading term $c_{0} = 744$ in the expansion \rf{largej2}; as a consequence, it comes with a doubly exponential dependence on the dilaton. In contrast, in the case \rf{beta2} with stabilization, the axion potential appears already in the second term of the single exponential approximation:  
 \be
  V_\beta^{j-\bar j}  = V_0 \left( 1 - \frac{\ln \beta}{2\pi} e^{- \sqrt{2\over 3\alpha} \varphi} + \frac{\ln \beta}{8 \pi^2} e^{-2\sqrt{2\over 3\alpha} \varphi} \ln(1 + 4 A  \sin^2(2 \pi \theta) )+ \ldots \right) \,.  
 \label{beta3}
 \ee
Indeed, this leads to the axion stabilization at $\theta = k/2$.

%%%%%%%%%%%%%%%%%%%%%%%%%%%%%%%%%%%%%%%
%%%%%%%%%%%
%% FIGURE 1: Scalar potentials with various A's

\begin{figure}[t!]
\vskip -10pt
\centering
\subcaptionbox{ $A=0$.\label{fig:ff2a}}{\includegraphics[scale=0.33]{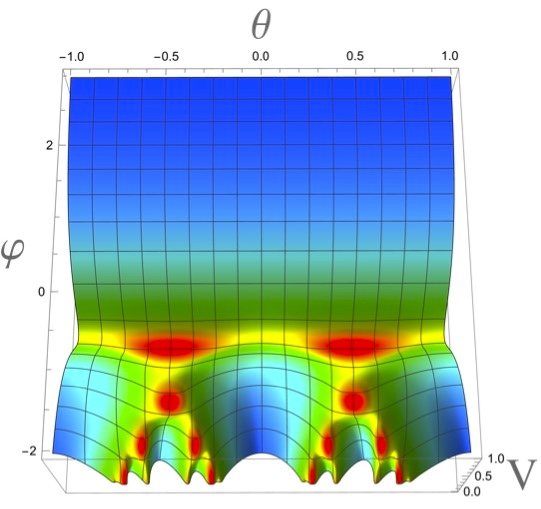}} 
\subcaptionbox{  $A=1$.\label{fig:ff2b}}{ \includegraphics[scale=0.33]{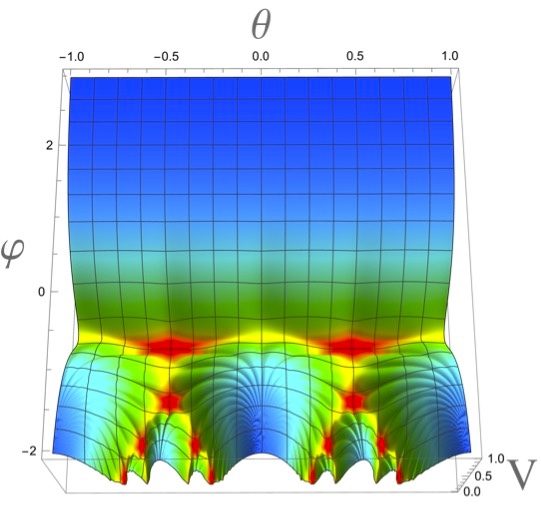}}
\subcaptionbox{ $A=10$.\label{fig:ff2c}}{ \includegraphics[scale=0.33]{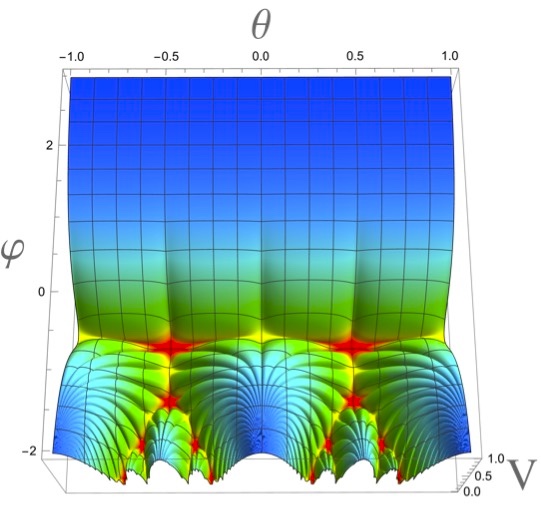}}
\subcaptionbox{ $A=100$.\label{fig:ff2d}}{ \includegraphics[scale=0.33]{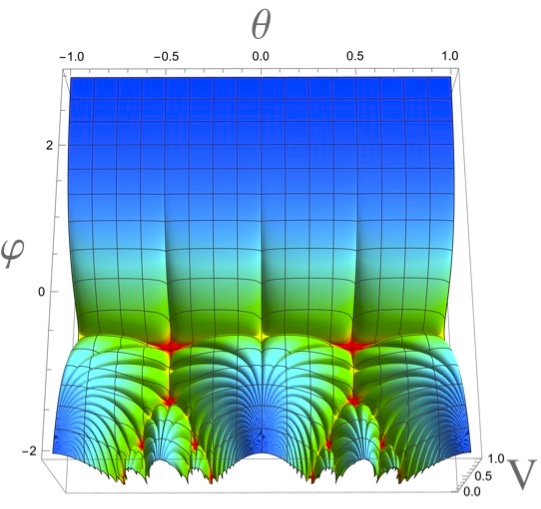}}
\vskip  10pt
\caption{\footnotesize The scalar potential \rf{beta2} for various A as a function of $\vp$ and $\theta$ for $3\alpha=1$, $\beta = 12^{3}$.  Setting $A = 0$, as in the upper left figure \ref{fig:ff2a},  coincides with the potential \rf{beta1} introduced in \cite{Kallosh:2024ymt}. Red areas on the figures show  Minkowski minima with V= 0.  Note that there are two such minima  at $\theta = \pm 0.5$, see App.~B.}
\label{ff2}
\end{figure}

%%%%%%%%%%%%%%%%%%%%%%%%%%

The plots of this potential for various values of the parameter $A$ are shown in Fig. \ref{ff2}. At $A = 0$, the upper part of the potential is almost exactly flat. At $A\not = 0$, the potential has a set of axion valleys at $\theta =  k/2$, and a set of ridges at $\theta =1/4+k/2$. The valleys and ridges, minima and maxima of the potential with respect to $\theta$, are almost invisible at the upper parts of all plots corresponding to large $\vp$ (as they appear at second order in the single exponential expansion). Nevertheless,  they strongly affect the axion dynamics, because their importance is actually exponentially enhanced. 

This can be seen from the equations of motion  for the homogeneous fields $\vp$ and $\theta$ in the hyperbolic geometry \eqref{hyperbolic}, given by \cite{Kallosh:2024whb}
 \begin{align}\label{eom}
\ddot\vp+3H\dot\vp+\sqrt{\frac{3\alpha}{2}}e^{-2\sqrt{\frac{2}{3\alpha}}\vp}\, \dot\theta^{2}+V_\vp=0 \ , \quad
\ddot\theta+3H\dot\theta-2\sqrt{\frac2{3\alpha}}\dot\vp\dot\theta+\frac{2}{3\alpha}e^{2\sqrt{\frac{2}{3\alpha}}\vp}V_\theta=0 \ .
\end{align}
At $\vp \gg \sqrt{\alpha}$ in the slow-roll approximation we have
\begin{align} 
3H\dot\vp = -V_\vp  \,, \qquad 
3H\dot\theta= -\frac{2}{3\alpha}e^{2\sqrt{\frac{2}{3\alpha}}\vp}V_\theta \,. \label{Es}
\end{align}
The full expressions for the potential gradients are rather lengthy; we will give only the results in the large $\vp$ limit, which are most important for understanding the inflation process. 
Moreover, to investigate inflation in this model in the large $\vp$ limit, we will assume that exp$(4\pi e^{\sqrt{2\over 3 \alpha}\vp}) \gg 1+4A$. This condition is typically satisfied until the end of inflation. Under these assumptions, the potential gradients take the form
\be
  V_{\vp} = \frac{V_0 \ln\beta }{\pi \sqrt{6\alpha} } e^{-\sqrt{\frac{2}{3\alpha}}\vp} \,, \qquad
  V_{\theta}  = \frac{2A V_0  \ln\beta}{\pi}  e^{-2\sqrt{2\over 3\alpha} \varphi} \frac{\sin(4 \pi \theta)}{1 + 4 A  \sin^2(2 \pi \theta)} \,.
\label{beta4}
\ee
At first sight, this would suggest that the dynamics would be dominated by the dilaton gradient. However, this does not include the effect of the hyperbolic geometry: the axion gradient is exponentially enhanced with a factor $e^{2\sqrt{\frac{2}{3\alpha}}\vp}$ in the equations of motion \eqref{eom} and their slow-roll approximation \eqref{Es}. This factor arises from purely geometric considerations: the proper distance \eqref{hyperbolic} between points with different $\theta$ values goes to zero high up on the plateau at large $\vp$. 

\begin{figure}[t!]
\centering
\subcaptionbox{$A = 10^{{-3}}$.\label{fig:f2a}}{\includegraphics[scale=0.43]{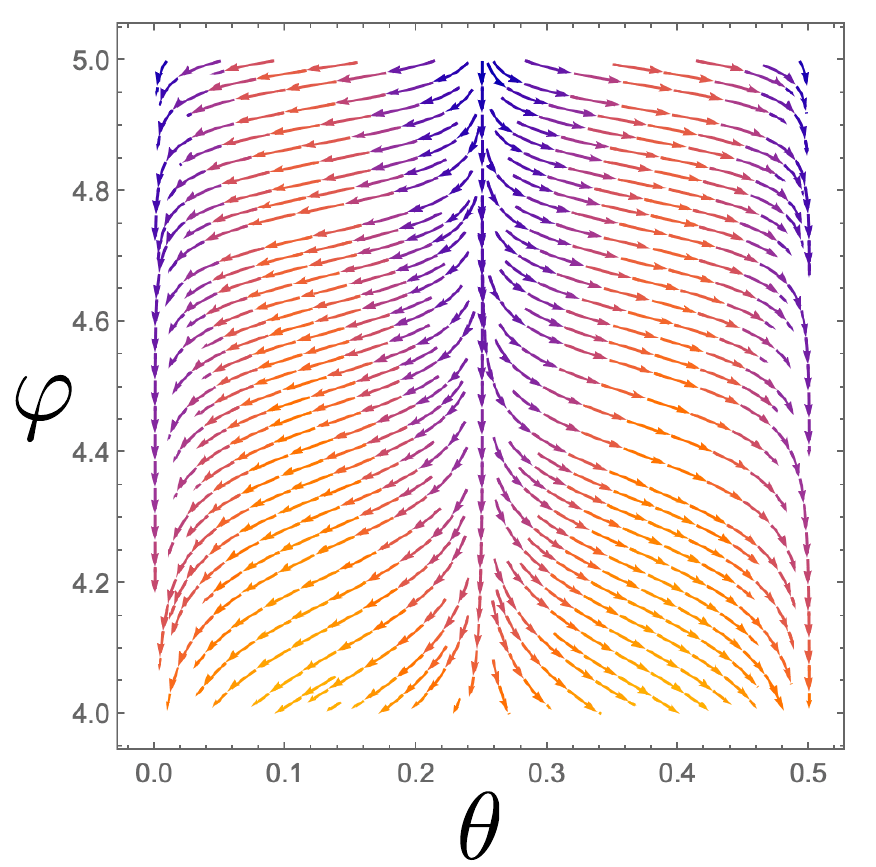}} 
\subcaptionbox{$A = 10^{{-2}}$.\label{fig:f2b}}{\includegraphics[scale=0.43]{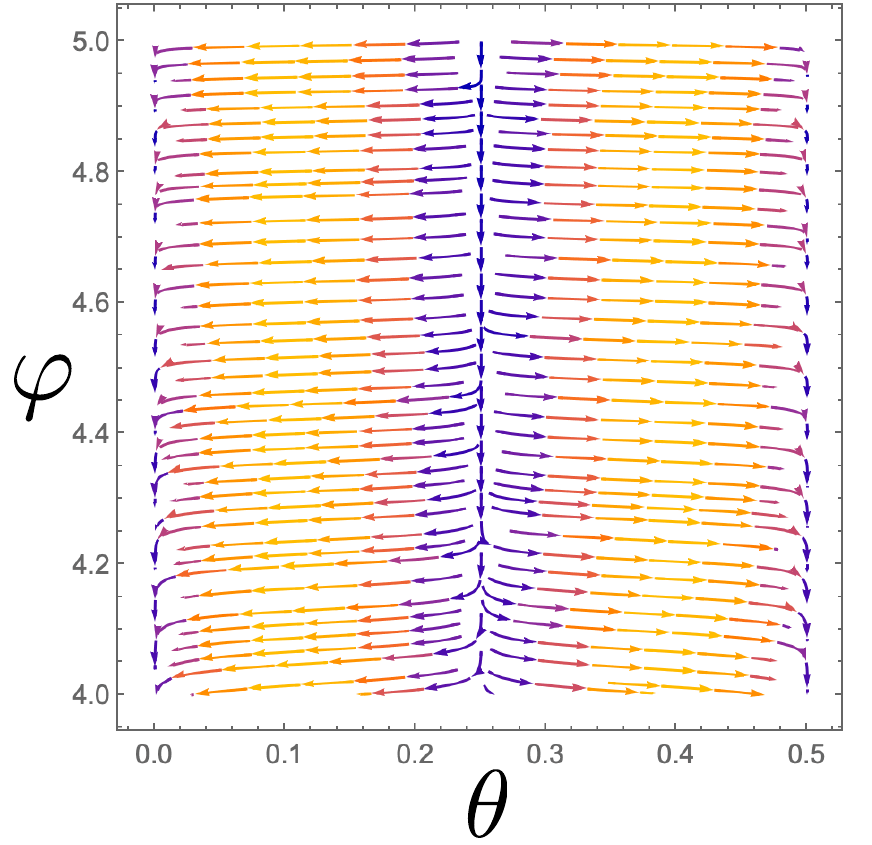}} 
\caption{\footnotesize {\it Left panel (\ref{fig:f2a})}: A stream plot of the slow-roll inflationary trajectories for $A = 10^{{-3}}$,  $3\alpha=1$ in the range $0<\theta < 1/2$, $5< \vp < 6$. Even the trajectories beginning very close to the ridge at $\theta = 1/4$  gradually reach the axion valleys at $\theta = 0$ or $\theta = 1/2$. This process is even more efficient for the trajectories at greater values of $\vp$ and $A$. {\it Right panel (\ref{fig:f2b})}: The same stream plot, for $A = 10^{{-2}}$. }
\label{f2}
\end{figure}

Taking this geometric factor into account, a 
comparison of the slow-roll equations \eqref{Es} shows that the relation between velocities  $\dot \vp$ and $\dot \theta$ is given by 
\be
{\dot\theta \over \dot\vp} =  {-{2\over 3\alpha} e^{2\sqrt{2\over 3\alpha} \vp} \, V_{\theta} \over  -V_{\vp}}    =  4\sqrt{2\over 3\alpha}\,  e^{ \sqrt{2\over 3 \alpha}\vp}\  {A   \sin 4\pi \theta\over     1+4A\, \sin^{2}2\pi \theta }\label{beta6} \ .
\ee
Therefore in the beginning of inflation at $\vp \gg 1$ one has $|\dot \theta| \gg |\dot \vp|$ for almost all initial values of $\theta$ (away from the ridges at $\theta = 1/4 + k/2$), unless $A$ is exponentially small. In other words, instead of moving down along the geodesic trajectories with constant $\theta$ as in the theory \rf{beta1}, at the beginning of the slow-roll inflation with  $\vp \gg 1$, the field $\theta $ in the theory \rf{beta2} rapidly rolls to the nearest ``axion valley''  $\theta   = {k \over 2}$. This is illustrated by a stream plot of inflationary trajectories beginning in the vicinity of the ridge of the potential at $\theta=1/4$ and then turning towards the axion valley, see Fig. \ref{f2}.

%%%%%%%%%%%%%%%%%%%%%%%%%%%%%%%%%%%%%%%
%%%%%%%%%%%
%% FIGURE 2: Slow roll trajectories A's

%%%%%%%%%%%%%%%%%%%%%%%%%%%%

Moreover, unless the parameter $A$ is very small, the motion of the field $\theta$ towards the axion valley at $\theta=0$ or $\theta=0.5$ can be even faster, violating the slow-roll regime for $\theta$. One can see this by calculating the slow roll parameter $\eta$, which is the smallest negative eigenvalue of the $2 \times 2$ matrix $\eta^I{}_J$ defined as 
\be
\eta^I{}_J = {g^{IK} D_K d_J V\over V} \ .
\ee
The only order-one  (i.e.~not exponentially suppressed) contribution to this matrix is $\eta^\theta{}_{\theta}$, which are large $\vp$ reads
\be
\eta^\theta{}_{\theta}  \approx {2\over 3 \alpha}\ e^{2\sqrt{\frac{2}{3\alpha}}\vp} {V_{;\theta \theta}\over V} =   {2\over 3 \alpha}\  {  2A(-2A +(1+2A) \cos 4\pi \theta ) \ln \beta^{2} \over  (1+ 4A\, \sin^{2} 2\pi \theta)^{2}}\ .
 \label{ratio}
\ee
In particular, along the axion valley at  $\theta = 0$ and at the ridge with at  $\theta =  1/4$, these take the values
\be
\eta^\theta{}_\theta(0) \approx  {  4A  \ln \beta^{2} \over  3\alpha} \ , \qquad
\eta^\theta{}_{\theta}(1/4) \approx  -{  4A  \ln \beta^{2} \over  3\alpha (1+4A)}  \ ,
\label{beta8}
\ee
clearly indicating the (in)stability of these axion values.

%%%%%%%%%%%%%%%%%%%%%%%%%%%%%%%%%%%%%%%
%%%%%%%%%%%
%% FIGURE 3: Slow roll parameter eta plots at various A's

\begin{figure}[t!]
\centering
\subcaptionbox{Slow roll parameter, $\eta^{\theta}_{\theta}$, for $A = 0.1$. \label{fig:etaA}}{\includegraphics[scale=0.43]{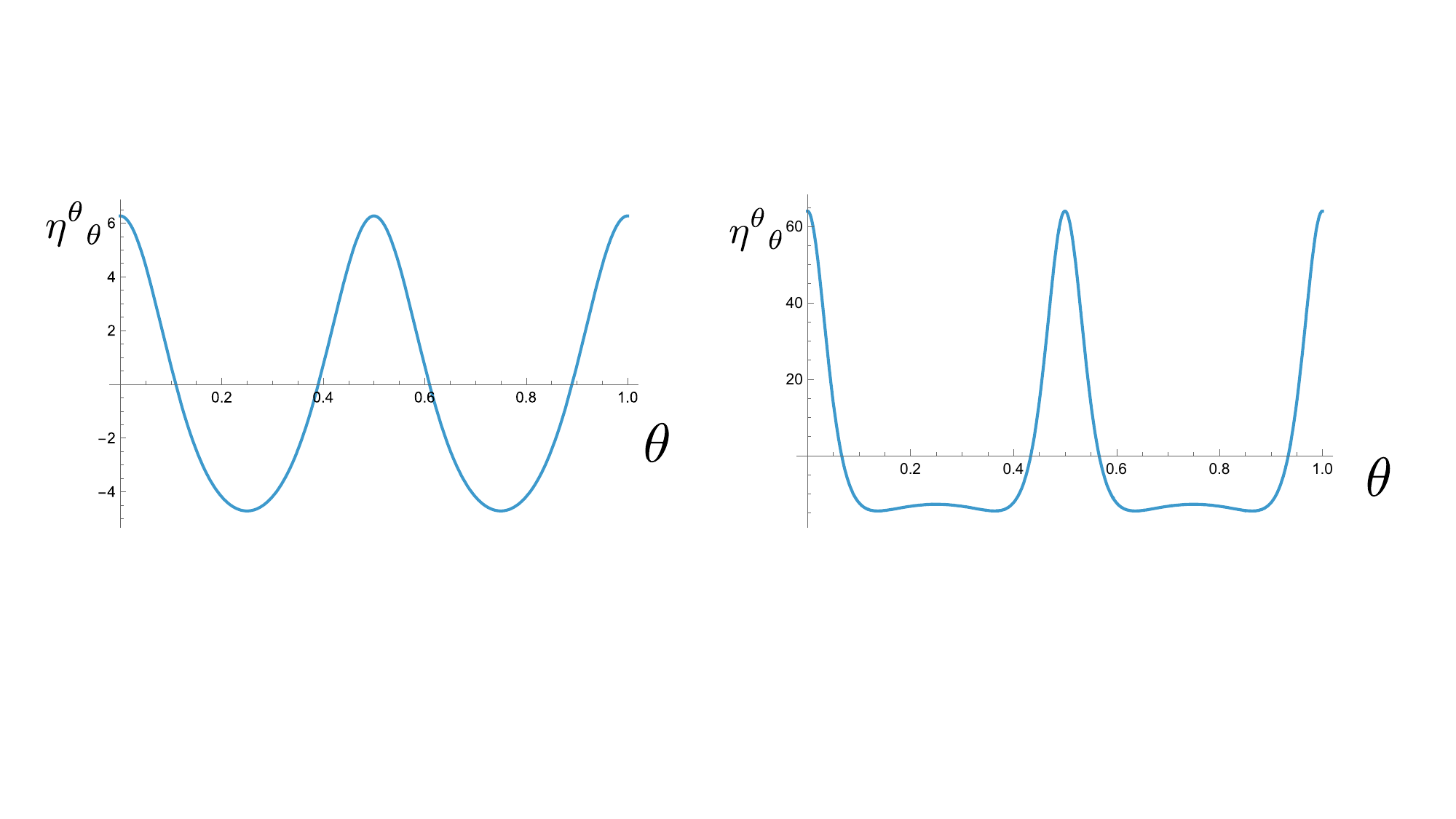}} 
\subcaptionbox{$A = 1$.\label{fig:etaB}}{\includegraphics[scale=0.43]{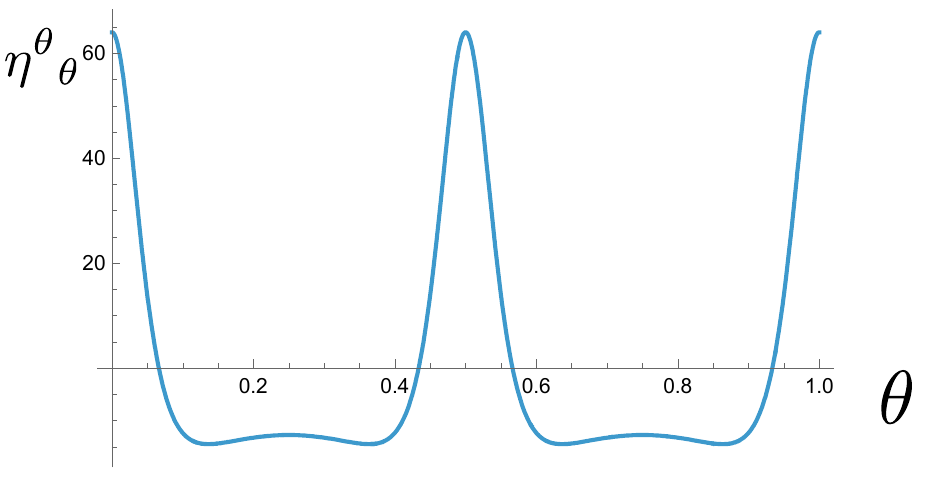}} 
\caption{\footnotesize  The slow-roll parameter $\eta^{\theta}{}_{\theta}$ \rf{ratio}  at $\vp=2$ for $3\alpha=1$,  $\beta = 12^{3}$.  In the left panel   $A = 1/10$, in the right one  $A = 1$. The  large   negative values of  $\eta^{\theta}{}_{\theta}$  at the ridge at $\theta = 1/4$ with $|\eta^{\theta}{}_{\theta}|\gg1$  explain why the axion field experiences a tachyonic instability. It   rapidly  falls to the nearest of the two axion valleys, $\theta = 0$ or $\theta = 1/2$, and remains strongly stabilized there since $\eta^{\theta}{}_{\theta}$ at the valleys is positive and big.  The same happens for all other values of $A \gtrsim 0.1$.}
\label{etatheta}
\end{figure}

%%%%%%%%%%%%%%%%%%%%%%%%%%%

Using the same parameters $\beta = 12^{3}$ and $\alpha = 1/3$ as in our other estimates in this paper, we find that the slow-roll regime with respect to the axion field breaks down already at $A \geq 0.02$, and this breaking becomes increasingly stronger at larger values of $A$, see Fig. \ref{etatheta}. This means that for $A \gtrsim 0.02$, the axion field experiences a strong tachyonic instability in the vicinity of the ridges at $\theta = 1/4 + k/2$, rapidly falls down to one of the axion valleys at $\theta = k/2$, and remains strongly stabilized theree,    see Fig. \ref{Middle}.  

Once the axion is stabilized, the difference between $ j(\tau)$ and $\overline {j(\tau)}$ disappears, and the potential along the $\theta = 1/2 +k$ valleys coincides with the original potential \rf{beta1}. However, in this new regime, the axion field in the axion valleys is heavy, $\eta^\theta{}_\theta(0) > 1$. Therefore, no isocurvature perturbations are generated in this scenario.{\footnote{ Note that after the field $\theta$ falls to the axion valleys at $A \geq 0.02$ we effectively have a single field inflation along the geodisics because the field $\theta$ is heavy and stabilized at the axion valleys $\theta = k/2$. One can clearly see it in Fig. \ref{Middle}. This is different from multifield inflation models with non-geodesic/rapidly turning trajectories, where the ``no $\eta$-problem'' issue was described in  \cite{Chakraborty:2019dfh,Aragam:2021scu,Bhattacharya:2022fze}.}

A detailed theory of the subsequent cosmological evolution depends on the valley to which the axion field falls. We consider first initial values for the axion $\theta$ in between $0$ and $0.25$, i.e.~on the left-hand side of Fig.~\ref{f2}. With these initial conditions, the field $\theta$ rapidly falls to the axion valley at $\theta = 0$, and then the field $\vp$ drives inflation while slowly rolling down towards the dS saddle point at $\vp = \theta = 0$ (or $\tau = i$).

The point $\vp = \theta = 0$ is unstable, and the fields eventually fall to one of the two minima, $\theta = \pm 0.5$, $\vp \approx -0.101$ (or $\tau = e^{\pi i/3}$ or $e^{2 \pi i/3}$). This happens because of the tachyonic instability, just as in the Higgs model or in the hybrid inflation scenario. A detailed theory of this process can be found in \cite{Felder:2000hj,Felder:2001kt}. As pointed out in \cite{Kallosh:2024ymt}, in the model of the type of \rf{beta1}, this process is not inflationary unless $\beta$ is very close to $1$. This conclusion is valid in the model \rf{beta2} near the saddle point as well. Indeed, one can check that the slow-roll parameter 
$\eta^\theta{}_\theta$ is large and negative. Indeed, we find that  at the saddle point $\vp = \theta = 0$ in the  model with $\beta =12^3, A=0.1 $ one has $\eta^\theta{}_\theta\approx -79.23$.

Therefore, when the field $\vp$ rolls to the saddle point at $\vp = \theta = 0$, a non-inflationary process of spontaneous symmetry breaking occurs, bringing the fields down to one of the two minima at $\theta = \pm 0.5$, $\vp \approx -0.101$. Since the probability of falling to  $\theta =  0.5$ is equal to the probability of falling to $\theta = - 0.5$, the universe becomes equally divided into domains with $\theta =  \pm 0.5$ separated by domain walls with  $\theta =0$. This makes the parts of the universe emerging as a result of inflation with $0 < \theta < 0.25$ grossly inhomogeneous and unsuitable for life as we know it.

Fortunately, there is no such problem in the exponentially large parts of the universe with initial values in the complementary range between $0.25$ and $0.5$. As an illustration, we will show the behavior of the fields $\vp$ and $\theta$ during inflation beginning at $\vp = 4$ and $\theta = 0.27$ (close to the ridge at $\theta = 0.25$) for $A = 0.1$ and $3\alpha=1$,  $\beta = 12^{3}$, see Fig. \ref{Middle}.
As one can see, the field $\theta$ almost instantly moves towards the valley with $\theta= 0.5$ and oscillates there with a rapidly decreasing amplitude.  By solving the classical equations for different initial values of $\theta$, one can easily check that these results are $\theta$-independent in the specified range $0.25 < \theta < 0.5$.

%%%%%%%%%%%%%%%%%%%%%%%%%%%%%%%%%%%%%%%
%%%%%%%%%%%
%% FIGURE 4:  Example phi theta plots for a trajectory that quickly stabilizes.

\begin{figure}[t!]
\centering
\subcaptionbox{Example trajectory $\vp(t)$ for $A=0.1$. \label{fig:midPh}}{\includegraphics[scale=0.43]{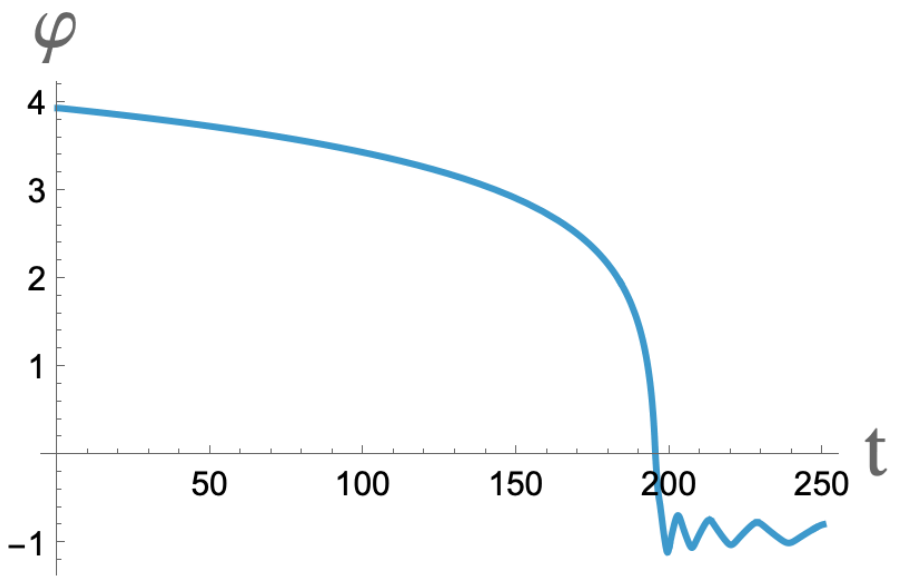}}  \quad
\subcaptionbox{Same trajectory, $\theta(t)$.\label{fig:midTh}}{\includegraphics[scale=0.43]{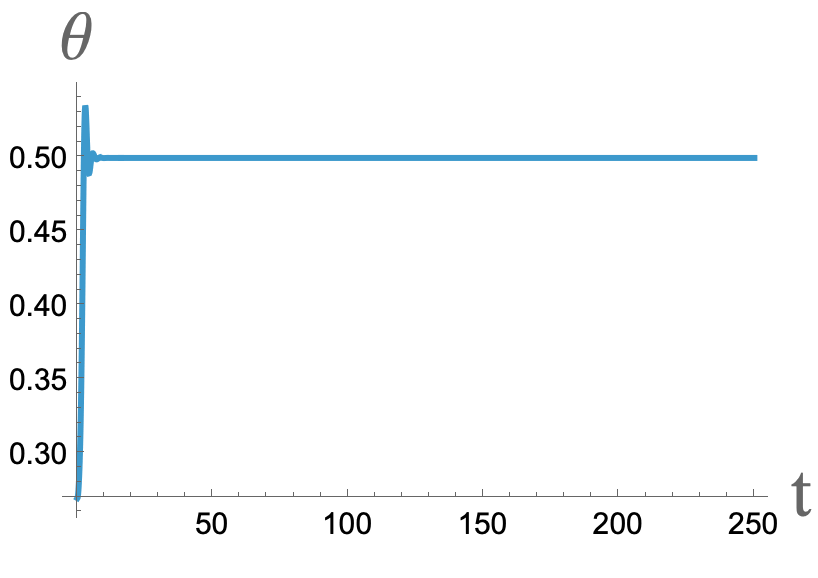}} 
\caption{\footnotesize The first plot on the left shows $\vp(t)$ for an example trajectory for, $A=0.1$, $\alpha=1$, and $\beta = 12^3$.  It starts at  $\vp(0)\approx 4$,  $\theta(0)=0.27$ and ends up after the exit from inflation in the second Minkowski minimum at $\vp(t)\approx -0.878$.  The second plot is of $\theta(t)$ for the same trajectory. which quickly reaches the valley at $\theta=0.5$, and then stays at the bottom of the valley while $\vp(t)$ reaches the minimum.  }
\label{Middle}
\end{figure}

%%%%%%%%%%%%%%%%%%%%%%%%%%%

It is instructive to make an analytic estimate of the time-dependent amplitude of the oscillations of the field $\theta$ about $\theta= 0.5$. According to \cite{Turner:1983he}, the energy density of the oscillating massive scalar field with mass $m \gg H$ decreases as the energy density of dust, ${m^{2}\theta^{2}\over 2} \propto a^{{-3}}$. During inflation, $a \sim e^{Ht}  \approx e^{N_{e}}$. Therefore the amplitude of the oscillations decreases as $a^{{-3/2}} \approx  e^{-3N_{e}/2}$. During the last  50 e-foldings, such oscillations decrease by a factor  $\sim e^{-75}$. This means that all strongly stabilized inflationary trajectories follow the axion valleys with exponentially large accuracy. 

After the field $\theta$ stabilizes at $\theta = 0.5$,  the field $\vp$ rolls down towards the first of the two red minima at $\theta= 0.5$ shown in Fig. \ref{ff2}. But because the barrier separating the two minima in this model is small, after falling to the first minimum at $\vp \approx -0.101$ the field $\vp$ does not stop there and rolls to the second minimum at $\vp \approx -0.878$. This is a very fast non-inflationary transition. Then, the field $\vp$ oscillates near the second minimum with a gradually decreasing amplitude. A more detailed description of these two Minkowski minima can be found in Appendix \ref{App:B}. 

Since no  isocurvature perturbations are generated in this scenario, inflation is described by the theory of a single field $\alpha$-attractors with the standard inflationary $\alpha$-attractor predictions
\be \label{pred2}
 A_{s} = {V_{0}\, N_{e}^{2}\over 18 \pi^{2 }\alpha} \ , \qquad n_{s} = 1-{2\over N_{e}} \ , \qquad r = {12\alpha\over N^{2}_{e}}  \ ,
\ee
with the universal dependence on the number of e-folds $N_e$.  These predictions provide a good fit to the Planck/BICEP/Keck constraints on $n_{s}$ and $r$ \cite{BICEP:2021xfz}.

 These results are valid in the standard single-field versions of $\alpha$-attractors, where the potential at its minimum is much smaller than the height of the plateau. This condition is satisfied for  $\beta = j[1] = 12^{3}$ which we discussed above. 

  However, as mentioned in \cite{Kallosh:2024ymt}, in the models with much smaller values of $\beta$,  the minimum of the inflaton potential along any fixed direction except $\theta = 0.5$ becomes considerably uplifted. The resulting potential shown in Fig. 6 of \cite{Kallosh:2024ymt} becomes very similar to the potential of the $\alpha$-attractor versions of hybrid inflation \cite{Kallosh:2022ggf}. As shown in \cite{Kallosh:2022ggf},  such uplift of the $\alpha$-attractor potential results in a controllable increase of the value of $n_{s}$, which may make $SL(2,\mathbb{Z})$ models compatible with the recent ACT result $n_{s} =0.9743 \pm  0.0034 $  \cite{Louis:2025tst}. For a recent discussion of this possibility in relation to the ACT results, see \cite{Kallosh:2025ijd}.

\section{$SL(2,\mathbb{Z})$ models with generalized axion stabilization}\label{general}

A generalized version of axion stabilization in these models is given by
\be
V =V_0\Big (1-{\ln \beta^{2}
\over  \ln\Big[|j(\tau)|^2+A\,
|e^{-2\pi i \gamma} j(\tau)  -e^{2\pi i \gamma}\ \overline {j(\tau)}|^{2}  + \beta^{2}\Big]}\Big ) \ .
\label{gamma}\ee
This potential is  $SL(2,\mathbb{Z})$ invariant because $ j(\tau)$ and $\overline {j(\tau)}$ are $SL(2,\mathbb{Z})$ invariants. It is also invariant under the transformation $\gamma \to  \gamma+ 1/2 $. At $\gamma = 0$, this potential coincides with the potential \rf{beta2} considered in the previous section.

At large $\vp$, this potential is given by
 \be
  V  = V_0 \left( 1 - \frac{\ln \beta }{2\pi} e^{- \sqrt{2\over 3\alpha} \varphi} + \frac{\ln \beta }{8 \pi^2} e^{-2\sqrt{2\over 3\alpha} \varphi} \ln\big(1 + 4 A  \sin^2(2 \pi (\theta + \gamma)) \big)+ \ldots \right) \,.  
 \label{large}
 \ee
In this model, the potential has the axion valleys at 
\be
\theta_{valley} ={k\over 2}  -\gamma   \ .
\ee 
For $\gamma =   k/2$, where $k\subseteq \mathbb{Z} $, one of the two axion valleys ends at a dS saddle point where $\vp=\theta=0$ (or $\tau=i$), just as in the model discussed in the previous section. However, for all other values of $\gamma$, the axion valleys do not end at the dS saddle point, and the field trajectories roll to one of the minima of the potential, avoiding the domain wall formation. In this context, the formation of domain walls is an exceptionally improbable outcome.

The most interesting example is provided by $\gamma = 1/4$. In this case, the potential \rf{gamma} acquires the most natural and symmetric form, where instead of $| j(\tau) -\overline {j(\tau)}|^{2}$ present in \rf{beta1} we have $| j(\tau) +\overline {j(\tau)}|^{2}$:
\be
V_\beta^{j+\bar j} =V_0\Big (1-{\ln \beta^{2}
\over  \ln\Big[|j(\tau)|^2+A\,
| j(\tau) +\overline {j(\tau)}|^{2} + \beta^{2}\Big]}\Big ) \ .
\label{beta2a}\ee
The full expression for the potential at large $\vp$ is therefore given by 
\be
  V_\beta^{j+\bar j}  = V_0 \left( 1 - \frac{\ln \beta }{2\pi} e^{- \sqrt{2\over 3\alpha} \varphi} + \frac{\ln \beta }{8 \pi^2} e^{-2\sqrt{2\over 3\alpha} \varphi} \ln(1 + 4 A  \cos^2(2 \pi \theta) + \ldots \right) \,.  
\label{beta2r}
\ee
This potential coincides with the potential \rf{beta3} up to the shift  $\theta \to \theta +1/4$. Therefore, an investigation of stabilization of the axion field with the potential \rf{beta2a}, \rf{beta2r} coincides with the corresponding investigation in the previous section, up to the shift of the position of the axion valleys by $\Delta \theta = 1/4$, see Fig. \ref{ff3b}.

%%%%%%%%%%%%%%%%%%%%%%%%%%%%%%%%%%%%%%%
%%%%%%%%%%%
%% FIGURE 5:  Scalar potential plot for A=100, gamma= 1/4.
\begin{figure}[t!]
\centering
\includegraphics[scale=0.14]{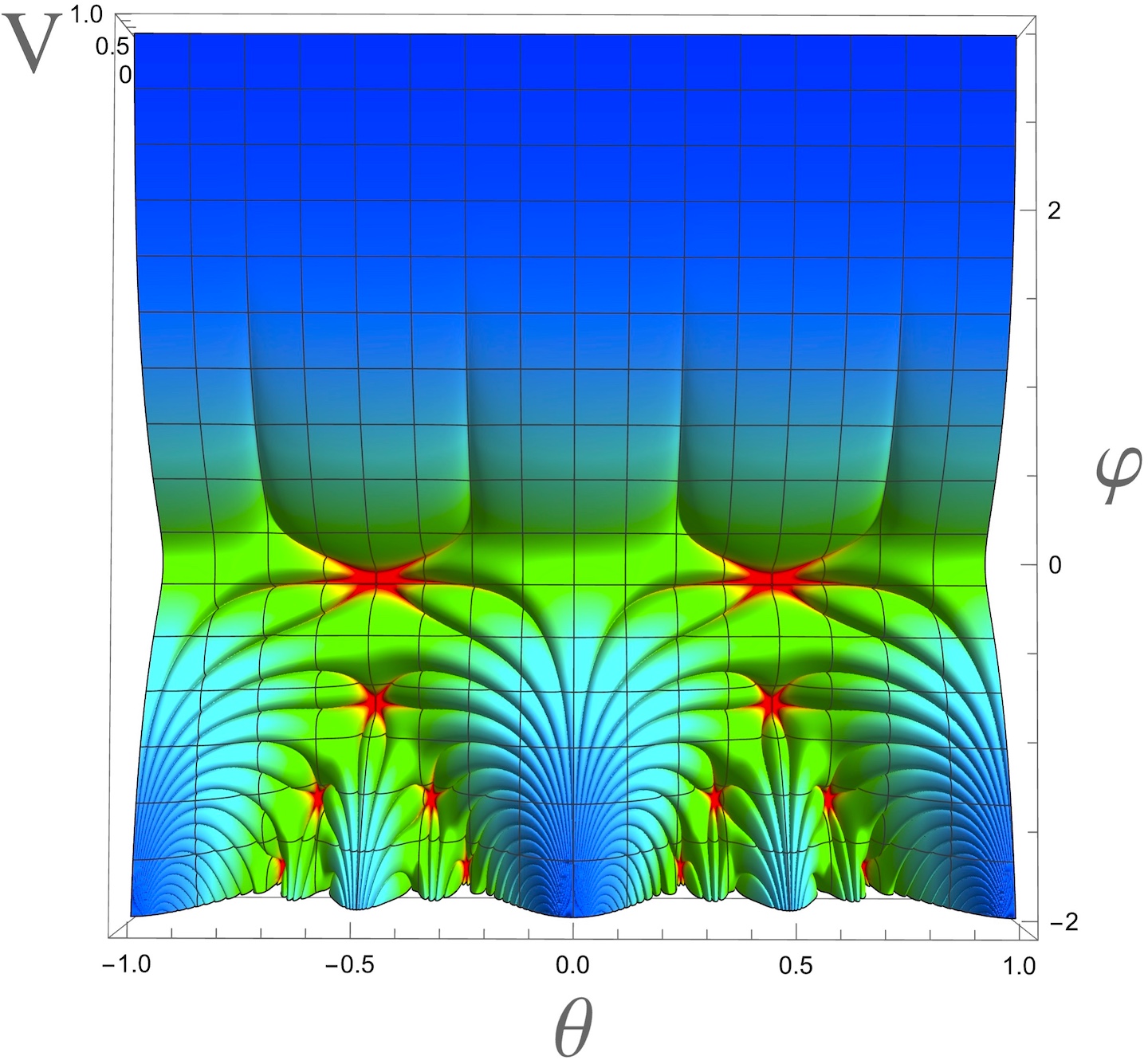} \vskip  10pt
\caption{\footnotesize The scalar potential \rf{beta2a} as a function of $\vp$ and $\theta$ for  $A = 100$ and $3\alpha=1$, $\beta = 12^{3}$. The stable inflationary trajectories are along the axion valleys at $\theta = 1/4 + k/2$; these are shallow and barely visible at large $\vp$, but their effect on the field dynamics is enhanced by the exponential term in \eqref{eom} and \eqref{Es}.}
\label{ff3}
\end{figure}

%%%%%%%%%%%%%%%%%%%%%%%%%%%

%%%%%%%%%%%%%%%%%%%%%%%%%%%%%%%%%%%%%%%
%%%%%%%%%%%
%% FIGURE 6:  Several trajectories stabilizing..
\begin{figure}[t!]
\centering
\includegraphics[scale=0.3]{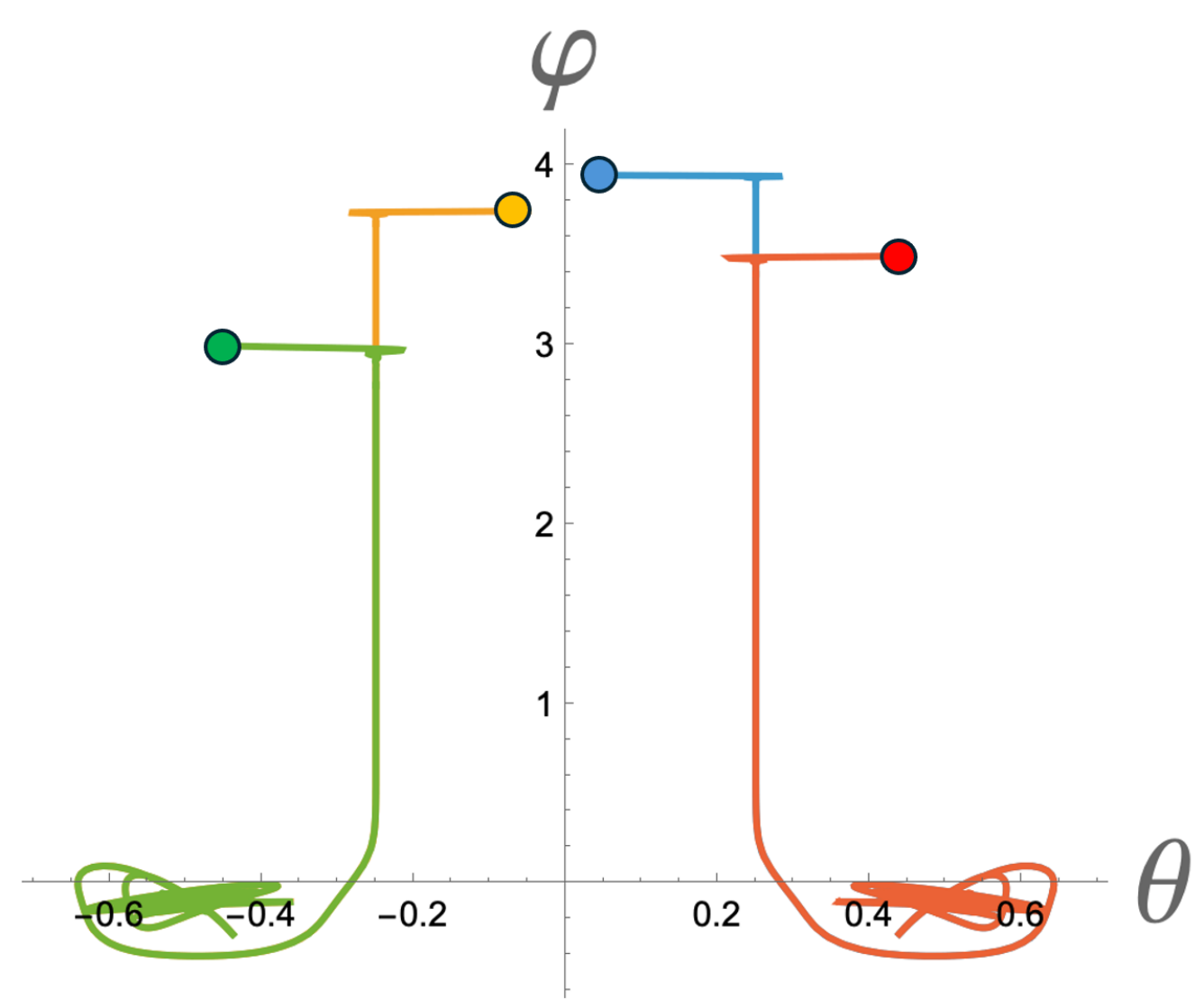} \vskip  10pt
\caption{\footnotesize Several inflationary trajectories in the theory \rf{beta2a} beginning at $-0.5 < \theta < 0.5$ for $A = 0.1$, and  $3\alpha=1$, $\beta = 12^{3}$. The small colored disks show the (static) initial conditions. As one can see, the field trajectories almost immediately reach the axion valleys at $\theta = \pm 1/4$. After that, the field $\vp$ rolls down, and all trajectories in each particular valley merge.   All trajectories reach the minima of the potential at $\vp \approx -0.101$, but not the second minimum at $\vp(t)\approx -0.878$.  The field $\theta$ does not change during inflation, no isocurvature perturbations are generated, and inflationary predictions are described by the theory of the single field $\alpha$-attractors \rf{pred2}. }
\label{ff3b}
\end{figure}

%%%%%%%%%%%%%%%%%%%%%%%%%%%

This shift of the position means that the trajectories confined to the axion valleys shown in Fig. \ref{ff3} differ from the previous case by $\Delta \theta = 1/4$, and hence do not arrive at the dS saddle point nor at $\theta=0.5$ just above the first minimum. These values led to a 50/50 division between domain wall formation or rolling through to the second minimum $\vp(t)\approx -0.878$, as discussed in the previous section. For the current case, with $\theta = 0.25 + k/2$, there is no domain wall problem, and the fields lose energy while they turn toward the Minkowski minimum. Due to this dissipation during this process, there is not enough energy to roll on to the second minimum, and all trajectories end up at the minima with $\vp \approx -0.101$. To illustrate this statement, we show inflationary trajectories in Fig. \ref{ff3b} for several different choices of initial conditions. In this model, one does not encounter the domain wall problem, and inflationary predictions coincide with the $\alpha$-attractor predictions \rf{pred2}.

Finally, we should note that in all previous calculations, we assumed that $A > 0$. That is why the axion valleys appear along the lines where the term $|j(\tau)  +  \overline {j(\tau)}|^{2}$ vanishes. However, the situation is different for $-1/4 < A<0$: in that case, the valleys appear along the lines when this term takes its maximal value. As a result, the plots of the potential \rf{beta2a} for $-1/4 < A<0$ start looking similar to the plots of the potential  \rf{beta2} for $A >0$, and vice versa. This (as well as the origin of the constraint $A > -1/4$) becomes especially clear from the comparison of the asymptotic expressions \rf{beta3} and \rf{beta2r}.

In the rest of the paper, we will continue the investigation of the models with $A>0$, but the results are easily extended to incorporate the models with $-1/4 < A<0$.

\section{Stabilized potentials in Killing coordinatess }\label{Sec:Kil}

As shown in \cite{Kallosh:2024ymt,Kallosh:2024pat}, the global structure of inflationary $SL(2,\mathbb{Z})$ potentials is highly nontrivial. Periodicity of the potential with respect to the axion field $\theta$ implies that the potential contains an infinite number of copies of the fundamental domain, see Fig. \ref{fig:CartesianA} for the potential \rf{beta2}.  Inflation can begin at the plateau with large values of $\vp$, at any value of $\theta$. Meanwhile,  at $\vp < 0$ the potential consists of infinitely many sharp ridges forming an intricate fractal structure. Naively, one would not expect that it is possible to have inflation at $\vp < 0$.

%%%%%%%%%%%%%%%%%%%%%%%%%%%%%%%%%%%%%%%
%%%%%%%%%%%
%% FIGURE 7:  Comparing \gamma =0 and \gamma=1/4 potentials in Cartesian coordinates.

\begin{figure}[t!]
\centering
\subcaptionbox{Potential \rf{gamma} with $\gamma=0$. \label{fig:CartesianA}}{\includegraphics[scale=0.55]{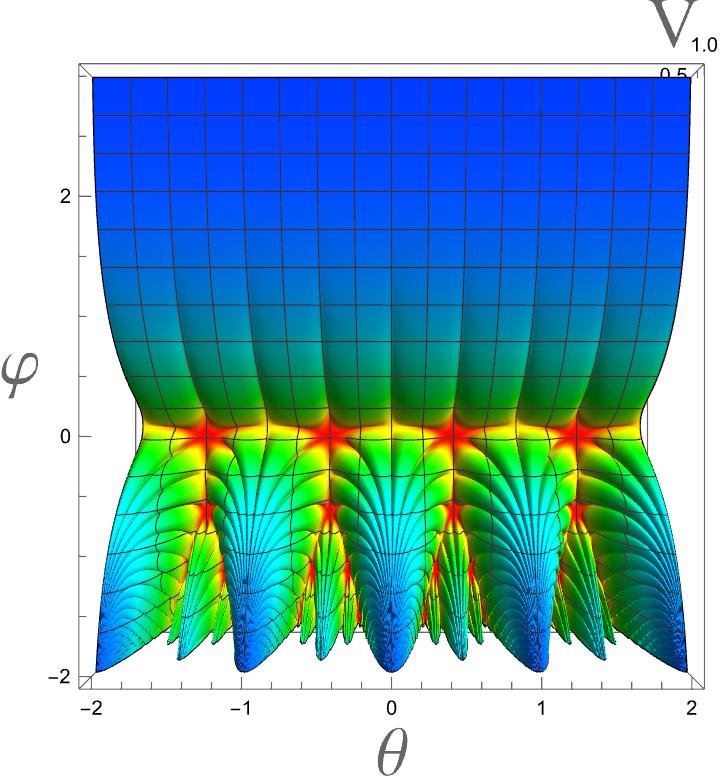}}  \qquad 
\subcaptionbox{Potential \rf{gamma} with $\gamma=1/4$. \label{fig:CartesianB}}{\includegraphics[scale=0.54]{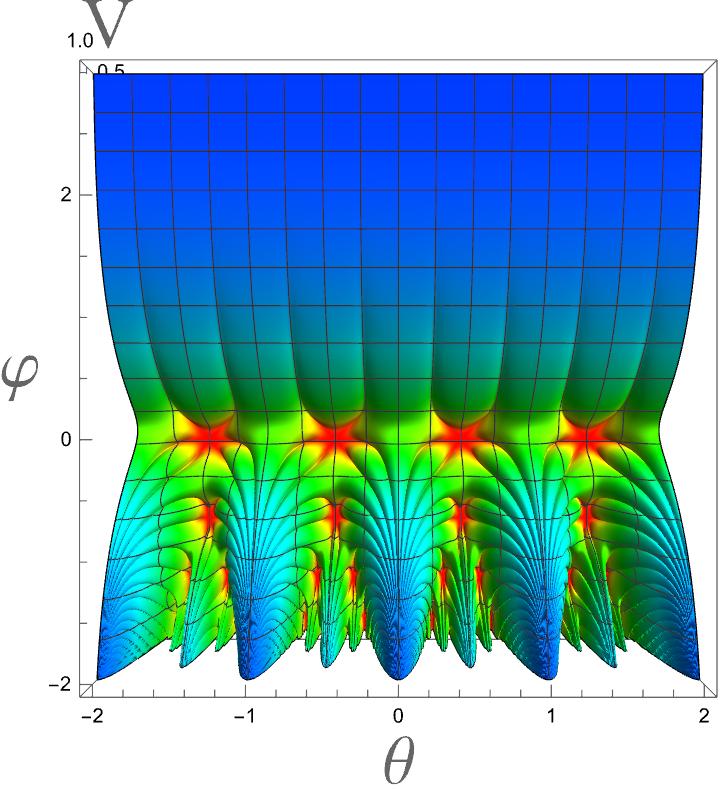}}
\caption{\footnotesize The left panel, \ref{fig:CartesianA}, shows the potential \rf{gamma}  for $\gamma=0$ in the Cartesian coordinates \rf{hyperbolic}.  This is equivalent to \rf{beta}.  The right panel, \ref{fig:CartesianB}, shows the potential for $\gamma=1/4$ and is thus equal to \rf{beta2a}.  Both potentials are shown for $A = 10$ and $3\alpha=1$, $\beta = 12^3$. On the left, valleys end in the  Minkowski minima or in the dS saddle points. On the right, all valleys end in the Minkowski minima.}
\label{Cartesian}
\end{figure}
%%%%%%%%%%%%%%%%%%%%%%%%%%%

However, the clear distinction between the (infinite number of) very sharp ridges downstairs and the very wide plateau upstairs is actually a mirage, an optical illusion created by hyperbolic geometry: in fact, as was shown in \cite{Kallosh:2024ymt,Kallosh:2024pat}, each of these ridges is physically equivalent to the entire upper part of the potential shown in Fig. \ref{Cartesian}. This means, in particular, that each of these ridges can support inflation just as well as the upper part of the half-plane shown in   Fig. \ref{Cartesian}. This statement is rather non-trivial and counterintuitive, raising many additional questions. For example, one may wonder what the inflationary trajectories may look like at $\vp <0$. As we will see, investigation of the potentials with the axion valleys provides a simple answer to this question.

We found it very useful to study the global structure of the $SL(2,\mathbb{Z})$ potentials  in Killing coordinates \cite{Carrasco:2015rva,Carrasco:2015pla,Kallosh:2024ymt,Kallosh:2024pat}  
\be\label{polar}
\tau =i  e^{\sqrt{2\over 3\alpha} (\tilde\vp-i \vartheta) } =  e^{\sqrt{2\over 3\alpha} \tilde\vp} \left(\sin\bigl(\sqrt{2\over 3\alpha} \vartheta\bigr) + i \cos\bigl(\sqrt{2\over 3\alpha} \vartheta\bigr) \right)\, , \quad -\pi/2 <  \sqrt{2\over 3\alpha} \vartheta < \pi/2\ .
\ee
 The hyperbolic geometry \eqref{hyperbolic} in these coordinates is 
\be
 ds^2 = 6 \alpha \, {d \tau d \bar \tau\over |\tau - \bar \tau|^2}= { (d \tilde\vp)^2+ (d \vt)^2\over \cos^2 (\sqrt{2\over 3\alpha} \vt)} \ .
\label{kin}\ee
The potential \rf{beta2} in Killing coordinates is shown in Fig. \ref{fig:KillingA}. The upper part of this figure, looking like a blue scallop shell, is an image of the entire upper part of the half-plane shown in Fig. \ref{fig:CartesianA}.  It shows the same periodic modulation that one can observe in Fig. \ref{fig:CartesianA}. This modulation shows the positions of the axion valleys. As we already mentioned, the axion can be effectively stabilized in these valleys even for $A \lesssim 1$, but to make this modulation more visible, we show it in Fig. \ref{Killing} for $A = 10$. 

%%%%%%%%%%%%%%%%%%%%%%%%%%%%%%%%%%%%%%%
%%%%%%%%%%%
%% FIGURE 8:  Comparing \gamma =0 and \gamma=1/4 potentials in Killing coordinates.

\begin{figure}[t!]
\centering
\subcaptionbox{Potential \rf{gamma} with $\gamma=0$. \label{fig:KillingA}}{\includegraphics[scale=0.145]{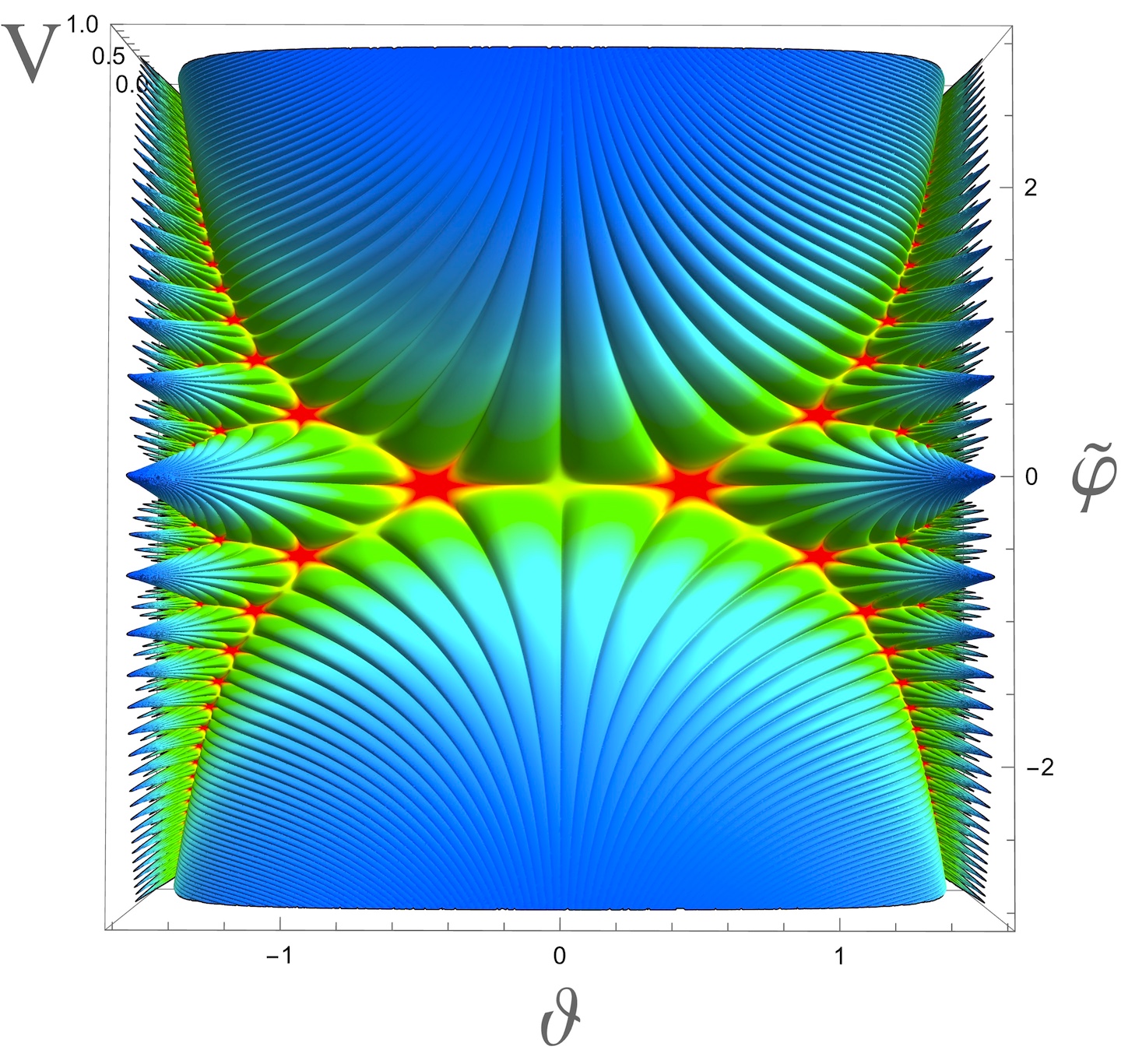}}  \quad 
\subcaptionbox{Potential \rf{gamma} with $\gamma=1/4$. \label{fig:KillingB}}{\includegraphics[scale=0.145]{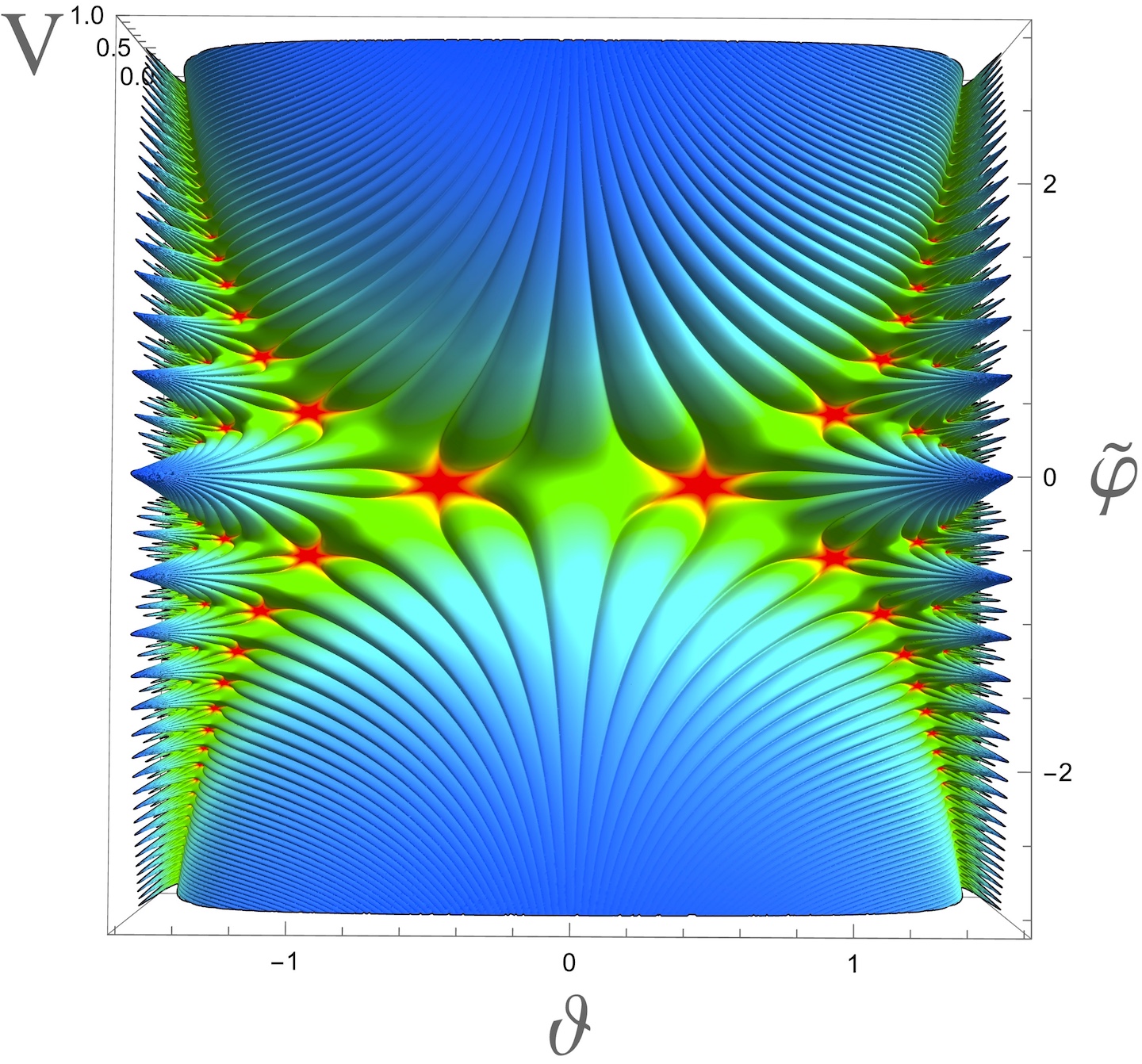}}
\caption{\footnotesize This figure shows the identical scalar potentials as in Figure \ref{Cartesian}, but now in Killing coordinates  \rf{polar}. The left panel, \ref{fig:KillingA}, shows the potential \rf{gamma}  for $\gamma=0$, equivalent to \rf{beta}, in Killing coordinates.  The right panel, \ref{fig:KillingB}, shows the potential for $\gamma=1/4$ and is thus equal to \rf{beta2a}.  Both potentials are shown for $A = 10$ and $3\alpha=1$, $\beta=12^3$. Again, on the left, valleys end in the  Minkowski minima or in the dS saddle points; on the right, all valleys end in the Minkowski minima.}
\label{Killing}
\end{figure}

%%%%%%%%%%%%%%%%%%%%%%%%%%%

The lower part of this figure also looks like a scallop shell. It is an image of the ridge at $\phi<0$, $\theta = 0$ shown in Fig. \ref{fig:CartesianA}. It looks like a mirror image of the shell in the upper part of Fig.  \ref{fig:KillingA}. Note that the potential  \rf{beta2}, as well as the metric \rf{kin}, is invariant with respect to the simultaneous change of variables $\tilde\vp \to -\tilde\vp$ and $\vt\to -\vt$. This means that the lower part of the potential shown in Fig. \ref{fig:KillingA} is physically equivalent to the upper part of this potential \cite{Kallosh:2024ymt,Kallosh:2024pat}.

Thus, inflation is indeed possible at the ridge  $\vp<0$, $\theta = 0$, and in fact at all other ridges shown in Fig. \ref{fig:KillingA}, see a detailed discussion of this issue in \cite{Kallosh:2024pat}. Moreover, now we can visually identify all stable inflationary trajectories: they correspond to all axion valleys going down from every ridge shown in Figs. ~\ref{fig:CartesianA} and  \ref{fig:KillingA}. These have a simple mathematical interpretation: they correspond to the lines where the function $j(\tau)$ takes real values, i.e., $ j(\tau)= \overline {j(\tau)}$.

One can perform a similar investigation of the potential \rf{beta2a}. In that case, the axion valleys, which are shown in the right panels of Figs. \ref{Cartesian} and \ref{Killing}, correspond to the lines where $j(\tau)$ is purely imaginary, i.e.  $ j(\tau)= - \overline {j(\tau)}$. All inflationary trajectories in this scenario end in the Minkowski minima with $V = 0$.

\section{Stabilized potentials in disk variables }\label{Sec:Disk}

Killing coordinates discussed in the previous section are very important for the investigation of the global structure of the $SL(2,\mathbb{Z})$ invariant potentials \cite{Kallosh:2024pat}. However, they do not allow us to show the entire potential in a finite size figure because the variable $\tilde\vp$ changes in the infinite range, $-\infty< \tilde\vp< +\infty$. Fortunately, one can circumvent this problem by using disk variables $Z$. They are related to the Cartesian coordinates $\tau$ of an infinite half-plane by a Cayley transform
\be\label{diskvar}
\tau = i {1+Z\over 1-Z}  = i {1+\rho \ e^{i\theta}\over 1-\rho \ e^{i\theta}}\ .
\ee
Here $Z = \rho \ e^{i\theta}$, $0\leq \rho  < 1$, $0\leq \theta \leq 2\pi$.
\

We hope to return to a discussion of the $SL(2,\mathbb{Z})$ invariant potentials in disk variables \rf {diskvar} in a separate publication. Here we will only show the plot of the potentials \rf{beta2} and \rf{beta2a}   in variables $\rho$ and $\theta$, see Fig. \ref{disk1000}. 

\begin{figure}[H]
\centering
\vskip -10pt
\includegraphics[scale=0.15]{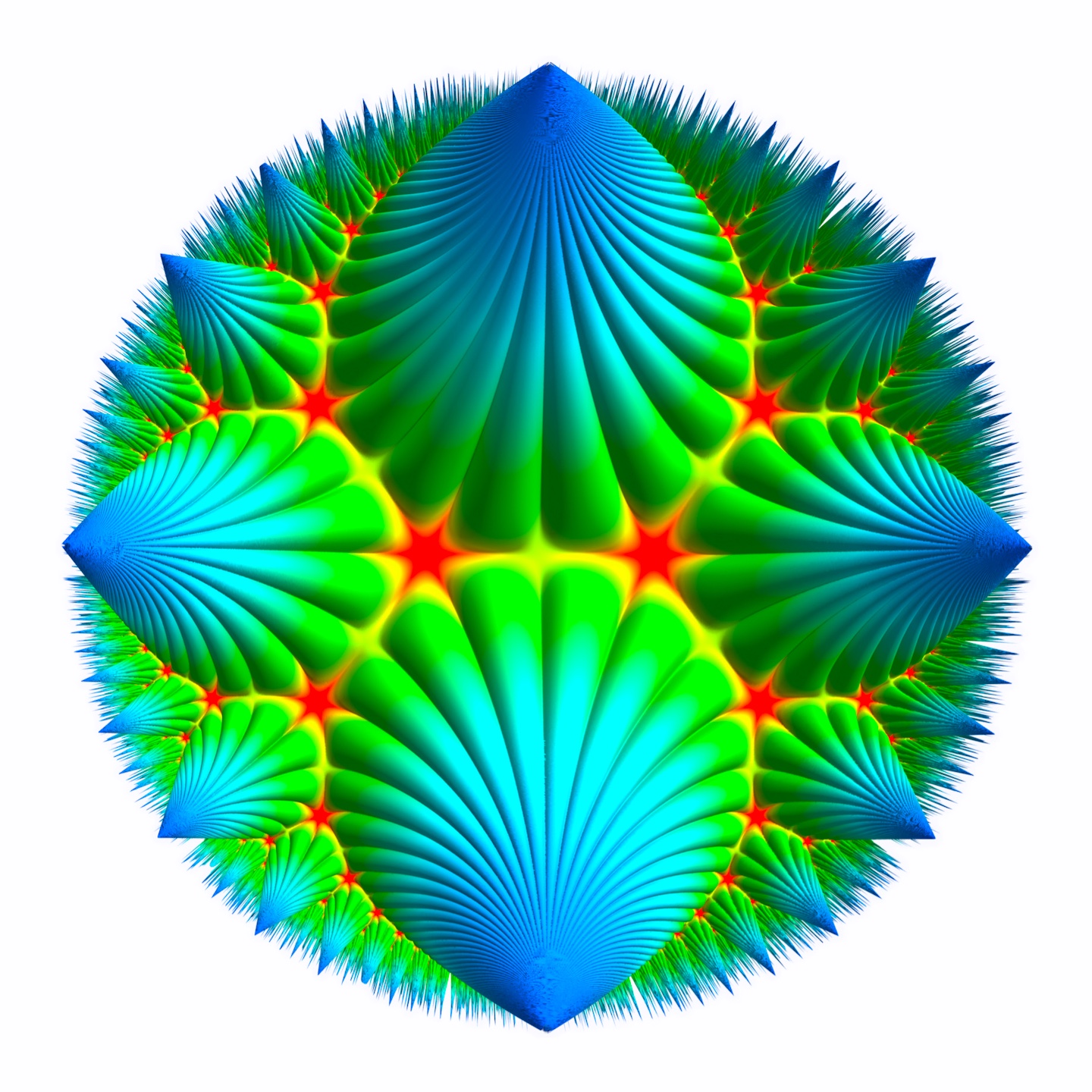}   \includegraphics[scale=0.15]{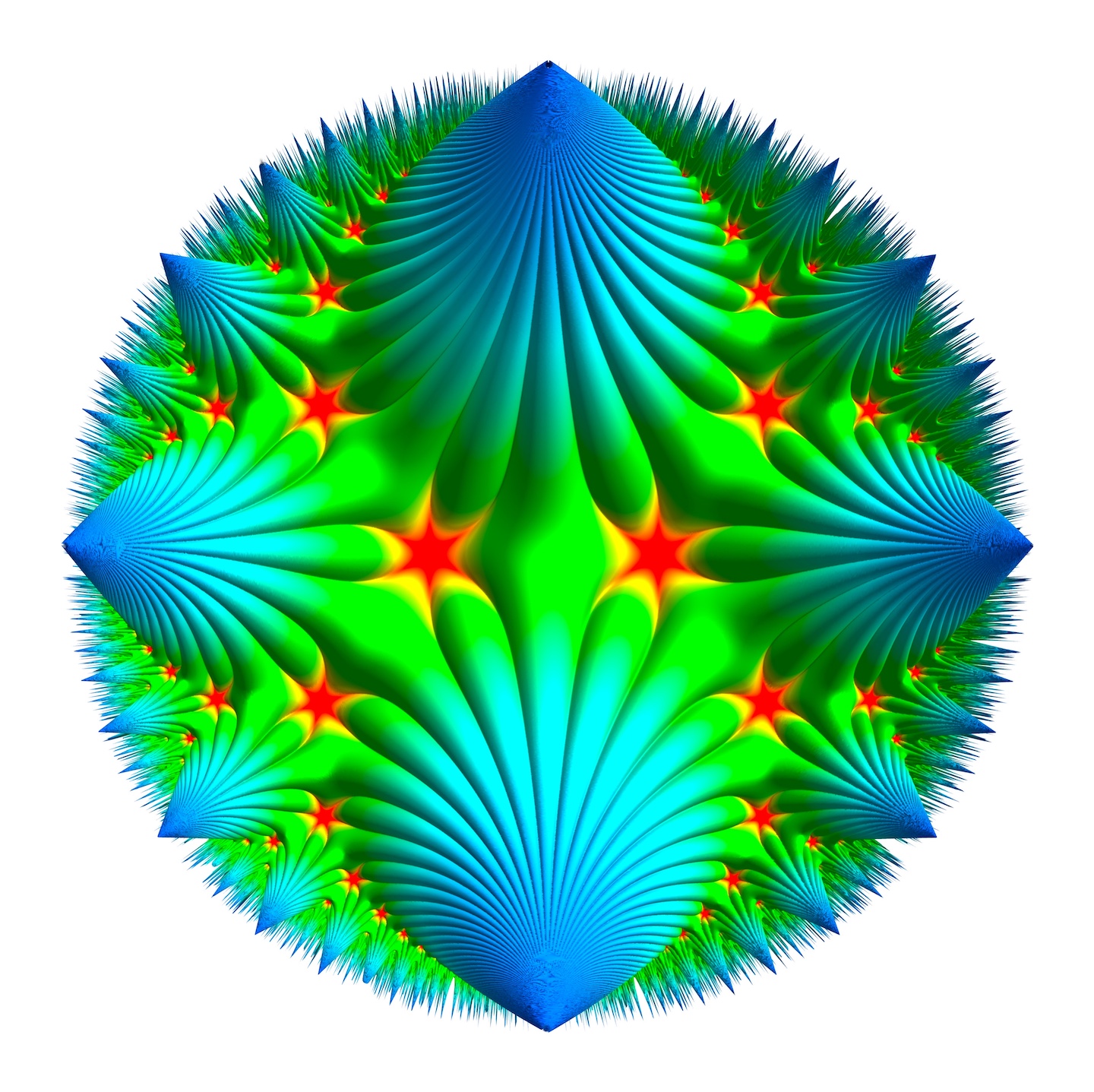}  
\caption{\footnotesize Potentials \rf{beta2} (left panel) and \rf{beta2a} (right panel) in disk variables $\rho$ and $\theta$, for $A = 10$ and $3\alpha=1$.  }
\label{disk1000}
\end{figure}

These potentials in disk variables look somewhat similar to the potentials in Killing coordinates shown in Fig. \ref{Killing}, but now they show the full space of all possible values of $\tau$ compressed to a disk of unit radius. The sharp peaks in this figure represent large inflationary plateaus, which are physically equivalent to each other. All stabilized inflationary trajectories in the models \rf{beta2} and \rf{beta2a} are shown by infinitely many axion valleys in the figures. 

Thus, the plots of the potentials with axion stabilization in disk variable instantly provide a detailed and complete map of the $SL(2,\mathbb{Z})$ invariant potentials,  including a map of all stabilized inflationary trajectories available in these theories.

\section{Summary}\label{Sec:sum}

As we already mentioned in the Introduction, the inflationary potentials in the simplest  $SL(2,\mathbb{Z})$ $\alpha$-attractor models in \cite{Kallosh:2024ymt} are almost exactly flat in the axion direction at large inflaton values \cite{Kallosh:2024whb}. This results in large isocurvature perturbations of the axion field. During inflation, these perturbations are harmless and do not feed into curvature perturbations. However,  these isocurvature perturbations may feed into the large-scale curvature perturbations {\it after inflation}  \cite{Kallosh:2024whb}. This interesting possibility deserves a detailed investigation and will be discussed in a separate publication \cite{6authors}. 

In this paper, we investigated the possibility of stabilizing the axion during inflation and avoiding the generation of isocurvature perturbations. This was a challenging task because all previously developed methods of stabilization of light fields during inflation in supergravity would violate $SL(2,\mathbb{Z})$ invariance. Therefore, the possibility to stabilize the axion during inflation without breaking $SL(2,\mathbb{Z})$  symmetry of modular inflation models is quite remarkable, especially since the mechanism of stabilization turned out to be very simple:  
It is sufficient to replace $|j(\tau)|^{2}$ in the potential \rf{beta1} by $|j(\tau)|^{2} + A | j(\tau) \pm \overline j(\tau)|^{2}$ as in \rf{beta2}, \rf{beta2a} or, more generally, by $|j(\tau)|^{2} +A\,
|j(\tau)\, e^{-i 2\pi \gamma} - \overline {j(\tau)} \, e^{i 2\pi \gamma}|^{2}$, where $A$ and $\gamma$ are  some constants, see eqs. \rf{gamma}.

\noindent There are two important features of stabilization in this scenario:
 \begin{itemize} 
\item First of all, the equation \rf{beta6} shows that, unless the stabilization parameter $A$ is exponentially small, all inflationary trajectories beginning at large values of the field $\vp$ quickly converge towards one of the axion valleys. This is qualitatively different from the situation in the models without stabilization ($A = 0$), where all initial values of $\theta$ appear equally probable.
 \item 
Secondly, the axion mass squared in these valleys becomes much greater than $H^{2}$ for $A \gtrsim 0.02$. No isocurvature inflationary perturbations of the field $\theta$ are produced in such models.
 \end{itemize}
As we discussed, in the model \rf{beta2} (with $\gamma=0$), for half of the initial conditions which lie in the range $0 < \theta < 0.25$, the inflationary trajectories end up at the saddle point at $\vp = \theta = 0$, which may lead to a domain wall problem. However, there is no such problem for the complementary and equally probable half with initial conditions $0.25 < \theta < 0.5$.  Moreover, this problem does not appear at all in the more general models of axion stabilization \rf{gamma}, \rf{beta2a} with parameter $\gamma$, as described in Section \ref{general}.

$SL(2, \mathbb{Z})$ cosmological models are inspired by string theory, with the concept of ``target space modular invariance" introduced in \cite{Ferrara:1989bc}. It was shown there that the corresponding $SL(2, \mathbb{Z})$ symmetry models of supergravity have an $SL(2,\mathbb{R})$ invariant kinetic term of the form \eqref{hyperbolic} where $3\a$ is an integer.  Similarly, the cases in \cite{Ferrara:2016fwe,Kallosh:2017ced} had seven discrete values of $3\a$. We thus conclude that the generic outcome of stabilized $SL(2, \mathbb{Z})$ models are the universal predictions of single-field $\alpha$-attractors attractors \rf{pred2}; moreover, due to the integer nature of $3\a$, it supports the seven Poincar\'e disk targets between $r\sim 1.6 \cdot 10^{-3}$ and $r\sim 1.1\cdot 10^{-2}$ (for $N_{e}$ = 50) for the LiteBIRD collaboration, as shown in Fig. 2 in \cite{LiteBIRD:2022cnt}. 

\newpage

\section*{Acknowledgement}
We are grateful to A. Achucarro, M. Braglia, E. Copeland, D. Lust, D. Wands, T. Wrase, and Y. Yamada for their helpful comments.
RK and AL are supported by SITP and by the US National Science Foundation Grant   PHY-2310429. JJMC is supported in part by the DOE under contract DE-SC0015910. DR would like to thank the SITP for the hospitality during the course of this work. 

\appendix  

 \section{ Supergravity  $SL(2,\mathbb{Z})$ models with  stabilized  axion}\label{App:A} 
 
 In supergravity, scalars are complex fields. Therefore, there was always a problem:  how to construct supergravity models that support single-field inflation? 

One of the simplest solutions was proposed in chaotic inflation in supergravity in \cite{Kawasaki:2000yn,Kallosh:2010ug,Kallosh:2010xz}, adding to the inflaton superfield $T$ a ``stabilizer'' superfield and using it to stabilize one of the fields in the inflaton superfield. The additional field can play the role of a heavy ``stabilizer'' superfield $S$ by allowing for stabilization via the 
bisectional curvature of the moduli space \cite{Kallosh:2010xz}. Alternatively, it can be a nilpotent ``stabilizer'' superfield $X$ with $X^2=0$ as developed in \cite{Carrasco:2015uma,Carrasco:2015rva,Kallosh:2017wnt}. The main idea there was to add to the \K\, potential a term of the form 
 \be
\Delta K=  A (T -\bar T)^2  F(T, \bar T) X\bar X \ ,
\label{bisec} \ee
 where $A$ is a constant, and $F(T, \bar T)$ is some function of the inflaton superfield $T$.
Therefore the bisectional curvature $R_{T\bar T  X \bar X}\neq 0$ is non-vanishing 
 during inflation. Inflation takes place at the $T=\bar T$ valley,  where the axion from the inflaton superfield is stabilized to zero.
 
The above stabilization terms, however, would break $SL(2,\mathbb{Z})$ invariance of the potential. That is why stabilization discussed in this paper is achieved by a modification of the inflationary potential preserving $SL(2,\mathbb{Z})$ invariance, following \cite{Kallosh:2024ymt}.  For a half-plane variable $T=-i \tau$,  the \K\, potential and superpotential are 
\begin{align}
K=&-3\alpha\ln(T+\bar T)+G_{X\bar X}(T, \bar T) X\bar X \, , \qquad W= W_0+F_X X \,,
\label{KW}\end{align} 
which can be packaged equivalently in the K\"ahler invariant function
\begin{align}
\mathcal G=&-3\alpha\ln(T+\bar T)+G_{X\bar X}(T, \bar T) X\bar X +\ln |W_0+F_X \, X|^2 \ .
\label{G}\end{align}
In the above, $W_0$ is a constant defining the mass of gravitino, $F_X$ is a constant that defines the auxiliary field vev, and
 \begin{align}
  G_{X\bar X}(T, \bar T)=\frac{|F_X|^2}{(T+\bar T)^{3\alpha}[ \Lambda  +V(T,\bar T)]+3|W_0|^2(1-\alpha)}\, , \qquad \Lambda =F_X^2 - 3W_0^2 \ .
 \label{GXX}
 \end{align}
This construction was gradually developed in \cite{Achucarro:2017ing,Yamada:2018nsk,Kallosh:2022vha}, where it was still believed that it is consistent only at $\a\leq 1$. 
The generalization of this proof for all values of $\a$, based on the use of the unitary gauge for local supersymmetry of supergravity, is to appear in {\it ``Half of the century of supergravity''} \cite{Kallosh:2025jsb}. 

The construction presented above was first given in \cite{Kallosh:2024ymt} and can be used for models where $V(T,\bar T)$ are $SL(2,\mathbb{Z})$ invariant potentials with Minkowski minima, as employed in \cite{Casas:2024jbw, Kallosh:2024ymt,Kallosh:2024pat,Kallosh:2024whb} as well as the stabilized potentials ones presented in this paper. The bosonic action following from this supersymmetric construction is\footnote{The construction in \rf{G}, \rf{GXX} gives a supersymmetric version of the potential $\Lambda +V(T, \bar T)$ for all values of  $T,  \bar T$, not constrained to inflationary trajectory with $T=\bar T$.}
\be
{ {\cal L} (T, \bar T)\over \sqrt{-g}} =  {R\over 2} - {3\alpha\over 4} \, {\partial T \partial \bar T\over ({\rm Re} \,  T )^2}- [\Lambda +V(T, \bar T)]  \ .
\label{hyper2}\ee
If $\Lambda =|F_X|^2 -3|W_0|^2>0$  and at the end of inflation $V(T, \bar T)=0$, the total potential in eq. \rf{hyper2} has a de Sitter vacuum. This action is 
$SL(2,\mathbb{Z})$ invariant if $V(T, \bar T)$ is $SL(2,\mathbb{Z})$ invariant.

\section{Two minima at each $\theta =  \pm1/2$}\label{App:B} 

In the main text, we have shown that our inflationary trajectories can end up at different minima of our potentials: see for example Figs.~\ref{Middle} and \ref{ff3b}, where inflation ends at up at either the second Minkowski minimum at $\vp(t)\approx -0.878$ or the first one $\vp(t)\approx - 0.101$. In both cases $\theta = \pm 0.5$. In this appendix, we will provide some details on these minima and their relation.

\begin{figure}[t!] 
\centering
\includegraphics[scale=0.6]{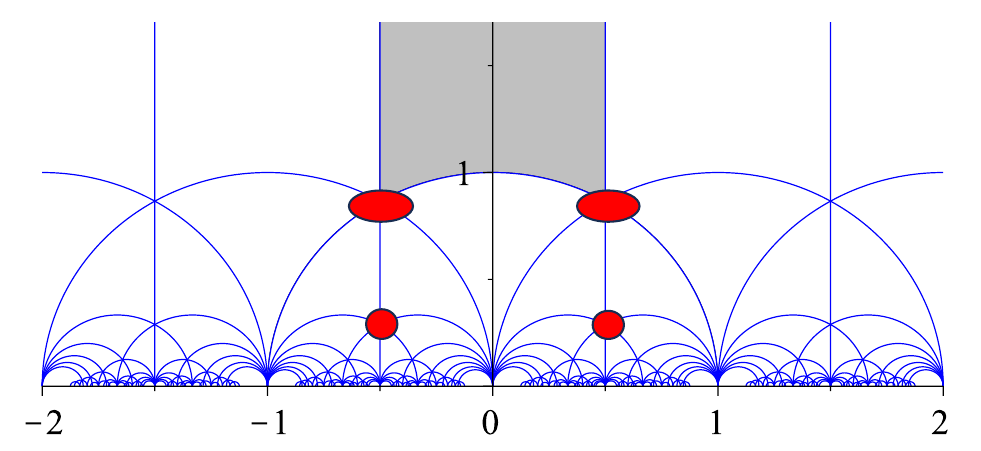} 
\caption{\footnotesize Tessellation of the hyperbolic  plane $\tau =\theta+iy$, $-\infty <\theta < \infty, \, y>0$ shows vertical bands which repeat each other. The grey area is the fundamental domain,  $-0.5 \leq \theta \leq 0.5$, $|\tau| \geq 1$, the other bands are shifted by 1 or -1 and repeated, see \href{https://en.wikipedia.org/wiki/Fundamental_domain}{Wikipedia}. We show here the positions of the two minima of our potentials, the first one, a red oval,  is at the boundary of the fundamental domain with $|\tau^{(1)}| =1$.  The second one, a red circle, is located at  $|\tau^{(2)}| =1/\sqrt{3}$ and hence does not border the fundamental domain.}
\label{2minima}
\end{figure}

The scalar potentials displayed in Figs. \ref{ff2} and \ref {ff3}  clearly show, in red, the first set of minima (of an elongated, elliptical shape) and the second set of minima (of circular shape), at every $\theta = 0.5 + k$. The same color coding for these two Minkowski minima is used in Fig.~\ref{2minima}, highlighting both minima and the fundamental domain as the grey area of the hyperbolic half-plane. For concreteness, we will focus here on the two minima at $\theta = -1/2$. The position of the first  minimum, located at the boundary of the fundamental domain, is given by
\be
\tau^{(1)} = e^{2\pi i\over 3}= -1/2 + i {\sqrt 3\over 2}\, , \qquad |\tau^{(1)}| =1 \ .
\ee
In terms of $\theta$ and $\vp$ defined via \eqref{hyperbolic} with $3\alpha$=1, this gives
\be
\theta^{(1)} = -1/2\ , \qquad \vp^{(1)} = {1\over \sqrt 2} \ln {\sqrt 3\over 2}\approx -0.101 \ .
\ee 
The second minimum is located at 
\be
\tau^{(2)} =  -1/2 + i \sqrt {1\over 12}\, , \qquad |\tau^{(2)}| =1/\sqrt{3} \ ,
\ee
and hence well outside of the fundamental domain; instead, it borders a copy of it.
Its location is
\be
\theta^{(2)}= -1/2 \ , \qquad \vp^{(2)}= {1\over \sqrt 2} \ln \sqrt {1\over 12}\approx -0.878 \ ,
\ee 
in terms of the axion and dilaton fields.

The two Minkowski minima are $SL(2,\mathbb{Z})$ images of each other, as can be seen in the following way. Employing the rules displayed in \cite{Kallosh:2024pat}, we can start with a particular initial value of $\tau_0$ and use the set of operations in eq. (6.1) in \cite{Kallosh:2024pat} to reach all possible $SL(2,\mathbb{Z})$  images. These operations are defined by a choice of integers $[q_0; q_1, \dots , q_n]$; in our case, the relevant choice turns out to be $[0; 2]$. This represents  the $SL(2,\mathbb{Z})$  operation $VU^{q_1} =V U^2$, where the corresponding matrix act as follows
\be
V U^2= \left(\begin{array}{cc}0 & -1 \\1 & 2\end{array}\right)  \ , \qquad  V U^2 \, \tau= -{1\over \tau +2}
\ee
One can check that the second minimum is the $SL(2,\mathbb{Z})$  image of the first minimum under this particular $SL(2,\mathbb{Z})$ operation:
\be
U^{-1}\,  V U^2 \, \tau^{(1)}=  \tau^{(2)} \ , \qquad 
\ee
It preserves $\theta^{(1)}=\theta^{(2)}=-1/2$ but moves $\vp^{(1)}\approx -0.101$ to $\vp^{(2)}\approx -0.878$, thus proving the equivalence between both minima. Interestingly, these are the only two minima at $\theta = -1/2$.

\bibliographystyle{JHEP}
\bibliography{lindekalloshrefs}
\end{document}